\documentclass[twocolumn,amsmath,amssymb,aps,preprintnumbers,showpacs,superscriptaddress]{revtex4}

\usepackage{graphicx}
\usepackage{dcolumn}
\usepackage{bm}

\newcommand{\ie}{{\it i.e. }}
\newcommand{\eg}{{\it e.g. }}
\newcommand{\cf}{{\it cf. }}

\newcommand{\bibdir}{bib/}      

\newcommand{\co}[2]{\ifcase #1 \or #2 \fi}

\newcommand{\lcmo}{{La$_{2/3}$Ca$_{1/3}$MnO$_3$}}

\begin{document}


\title{Laser microscopy of tunneling magnetoresistance
in manganite grain-boundary junctions}

\author{M.~Wagenknecht}
\author{H.~Eitel}%
\author{T.~Nachtrab}%
\affiliation{%
Physikalisches Institut --
Experimentalphysik II, Universit\"{a}t
T\"{u}bingen, Auf der Morgenstelle 14,
D-72076 T\"{u}bingen, Germany
}%
\author{J.~B.~Philipp}
\author{R.~Gross}
\affiliation{ Walther-Meissner-Institut,
Bayerische Akademie der Wissenschaften,
Walther-Meissner-Str. 8, D-85748
Garching, Germany
}%
\author{R.~Kleiner}
\author{D.~Koelle}
\email{koelle@uni-tuebingen.de}
\affiliation{%
Physikalisches Institut --
Experimentalphysik II, Universit\"{a}t
T\"{u}bingen, Auf der Morgenstelle 14,
D-72076 T\"{u}bingen, Germany
}%

\date{\today}

\begin{abstract}
Using low-temperature scanning laser
microscopy we directly image electric
transport in a magnetoresistive element,
a manganite thin film intersected by a
grain boundary (GB). Imaging at variable
temperature allows reconstruction and
comparison of the local resistance vs
temperature for both, the manganite film
and the GB. Imaging at low temperature
also shows that the GB switches between
different resistive states due to the
formation and growth of magnetic domains
along the GB. We observe different types
of domain wall growth; in most cases a
domain wall nucleates at one edge of the
bridge and then proceeds towards the
other edge.
\end{abstract}

\pacs{72.25.Mk, 75.47.Lx, 75.60.Ch, 75.70.-i}   

%
%
%

\maketitle


Doped perovskite manganites of
composition R$_{1-x}$A$_x$MnO$_3$ (R:
rare-earth, A: alkaline-earth element)
have attracted renewed interest over the
last years due to the interesting
interplay between charge, spin, orbital
and structural
degrees of freedom\cite{Coey99}. 
This is related to ordering phenomena which lead \eg
to the colossal magnetoresistance (CMR) effect\cite{Helmolt93,Jin94a} 
when a large magnetic field $B$ in the
Tesla-range is applied close to the Curie
temperature $T_C$ to align the magnetic
moments of the Mn ions, which increases
electric conductivity by up to several
orders of magnitude. Moreover, at
temperatures $T<T_C$ a substantial
magnetoresistance (MR) in much lower
fields ($\sim$10\,mT) has been found in
polycrystals and thin films with grain
boundaries (GBs)\cite{Hwang96,Gupta96,Mathur97}. 
Such a low-field MR effect allows
switching by small magnetic fields
between a low- and high-resistance state,
characterized by parallel and
antiparallel orientation of the
magnetization of two ferromagnetic grains
separated by a GB. This effect can be of
considerable interest for
applications like magnetic storage\cite{Prinz98}. 
However, a thorough understanding of the
magnetotransport properties in manganite
GBs is still lacking, and the models,
proposed  to describe those properties
are controversial. Whereas some suggest
activated carrier transport within a
defective region adjacent to the GB (with
depressed magnetic order)\cite{Evetts98}, 
or transport through an interfacial
depletion layer at an abrupt
metal-semiconductor contact at the
GB\cite{Gross00,Glaser02}, 
most of them are based on spin-polarized
tunneling\cite{Moodera99a} 
between ferromagnetic grains through an
insulating GB barrier\cite{Hwang96}, 
including \eg resonant\cite{Sun03} 
or inelastic\cite{Hoefener00} 
tunneling via localized states in the
barrier, which can be \eg due to
paramagnetic impurities\cite{Guinea98} 
that may be magnetically ordered\cite{Ziese99}. 
Most experiments were performed on
well-defined, individual GB junctions or
junction arrays fabricated by epitaxial
growth of manganite films on bicrystal
substrates\cite{Mathur97,Hoefener00,Steenbeck98,Philipp00,Blamire03,Gunnarsson04}. 
Large low-field MR ratios up to 300\,\% at $T$=4.2\,K
were found in
individual \lcmo (LCMO)
GB junctions\cite{Philipp00}. 
Based on the Julliere model for elastic
spin polarized tunneling\cite{Julliere75}, 
such a large tunnelling magnetoresistance
(TMR) effect can be achieved in materials
with large spin polarization $P$. In this
sense, the doped manganites are very
promising materials for device
applications due to their half-metallic
nature with $P$ close to unity\cite{Park98}. 
Accordingly, high TMR ratios have been
also found for trilayer spin valve
devices using doped manganites\cite{Sun96,Viret97,Jo00}. 

Obviously, the magnetotransport
properties of any TMR element are
affected by the magnetization of the
ferromagnetic electrodes close to the
barrier, and in particular by magnetic
domain formation and domain wall motion
during field switching. Therefore, a
better understanding of magnetotransport
in GB junctions from doped manganites
requires consideration
of magnetic domains\cite{Philipp00,Gunnarsson04,Todd99}. 
The measured TMR always represents an integral
quantity obtained by averaging over the area of the
tunnel junction. Both in view of applications and from
a fundamental point of view it is of high interest to
see how locally the TMR depends on the magnetic
properties of the electrodes and on the way the
magnetization is switched by a magnetic field. Here we
show that low-temperature scanning laser microscopy
(LTSLM) allows to locally detect the magnetoresistance
of LCMO  grain boundaries. Besides the more or less
homogeneous high- and low-resistance states we also
find intermediate states where low- and
high-resistance regions coexist. Our investigations
give insight into the magnetic domain formation near
the grain boundary and its impact on the
magnetoresistance of these devices.

\begin{figure}[tb]
\center{
\includegraphics[width=0.75\columnwidth,clip]
{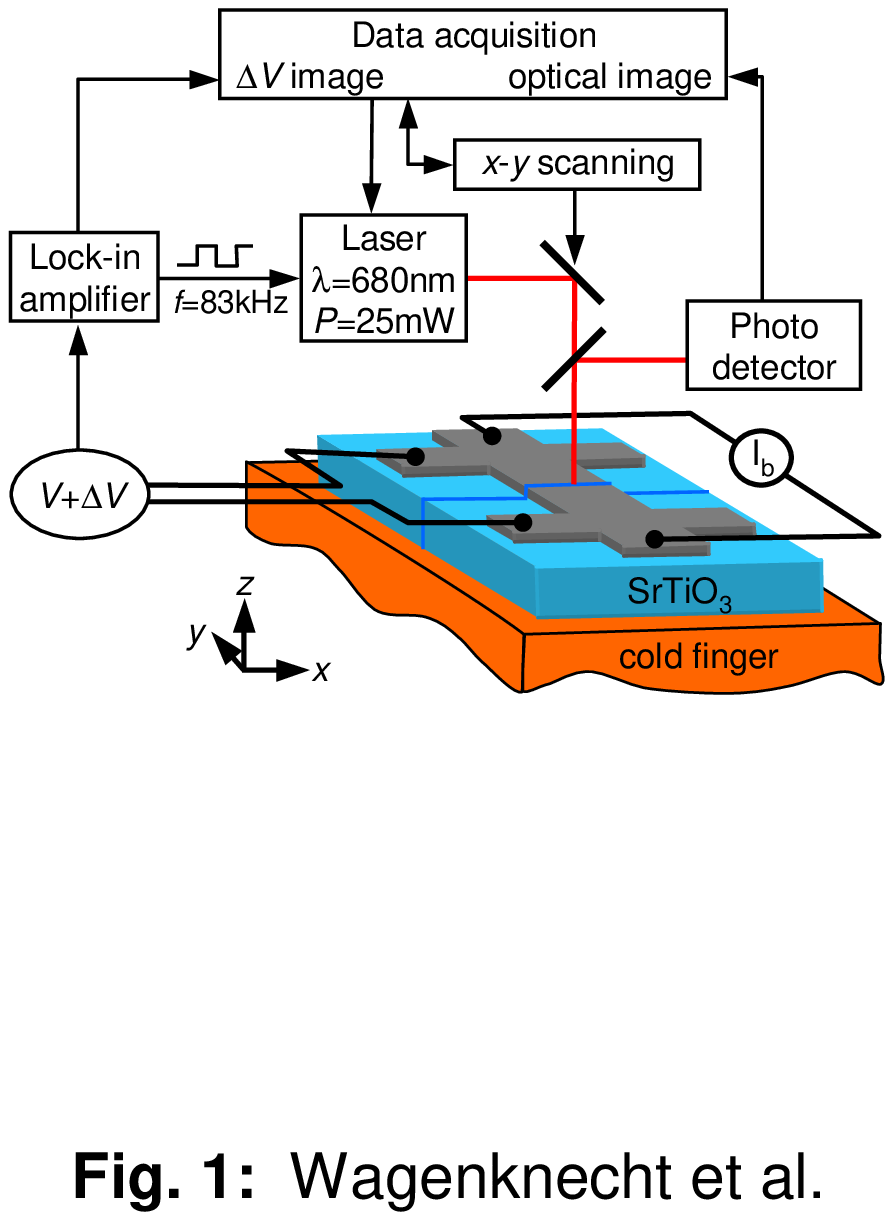}} %
\caption{(Color online) LTSLM setup: The
sample is mounted on a cold finger; all
other components are at 300\,K. The laser
beam is focused onto the LCMO microbridge
at position $(x_0,y_0)$ creating a hot
spot $(\delta T\approx 3\,\rm K$,
diam.~$\sim1.5\,\mu\rm m$). The local
change in sample resistance induces a
change $\Delta V$ of the globally
measured voltage $V$, creating the
electrical image $\Delta V(x_0,y_0)$. The
reflected beam intensity vs. $(x_0,y_0)$
produces an optical image. \label{fig:1}}
\end{figure}

To provide insight into the interplay
between electric transport and magnetic
domain formation, we studied GB junctions
made from 80nm thick LCMO thin films,
epitaxially grown by pulsed laser
deposition on symmetric [001] tilt
{SrTiO$_3$} bicrystals with $24^\circ$
misorientation angle\cite{Philipp00}. 
The films are patterned into
$W=30\,\mu\rm m$ wide bridges crossing
the GB. Voltage pads separated by
$L=70\,\mu\rm m$ along the bridge allow
4-point measurements. The samples are
mounted on a cold finger (at $5\,{\rm
K}\le T\le 300\,{\rm K}$) inside a $^4$He
optical flow cryostat. A Helmholtz coil
produces a magnetic field up to 35\,mT in
the ($x,y$) plane of the film, parallel
to the GB (along $x$) for measurements of
the integral resistance $R(T,B)$. For
imaging, the beam of a 25\,mW diode laser
($\lambda=680\,\rm nm$) is focused onto
the film surface [\cf Fig.\ref{fig:1}].
At the beam position $(x_0,y_0)$ the sample is locally
heated by $\delta T(x_0,y_0)\approx 3\,\rm K$ within
an area of diameter $d\approx 1.5\,\mu\rm m$,
approximately given by the laser spot size. This local
hot spot changes the local resistance $\sim (\partial
R/\partial T)|_{(x_0,y_0)}\cdot \delta T(x_0,y_0)$,
which induces a change of $R$ and consequently a
change $\Delta V(x_0,y_0)\propto I_b\cdot(\partial
R/\partial T)|_{(x_0,y_0)}\cdot\delta T(x_0,y_0)$ of
the total voltage $V$ across the current biased
bridge. Typically, $\Delta V$ is of order
1-20\,$\mu$V, for a bias current $I_b=5\,\mu\rm A$
(used for all measurements discussed below),
%
and much larger than the thermoelectric voltage change
$\Delta V_{th}$ ($I_b=0$) \cite{Shadrin98,Zhang02b} in
our samples.
%
To improve the signal-to-noise ratio, we blank the
laser beam at a frequency $f=83\,\rm kHz$ and detect
$\Delta V$ with a lock-in technique. An electrical
image is obtained by monitoring $\Delta V(x_0,y_0)$.
Simultaneously we detect the reflected beam intensity
vs. $(x_0,y_0)$ to obtain an optical image of the
scanned surface. We investigated two different LCMO
bridges intersected by a GB which showed qualitatively
the same behavior. Below we present data obtained on
one of those bridges.


\begin{figure}[b]
\center{
\includegraphics
{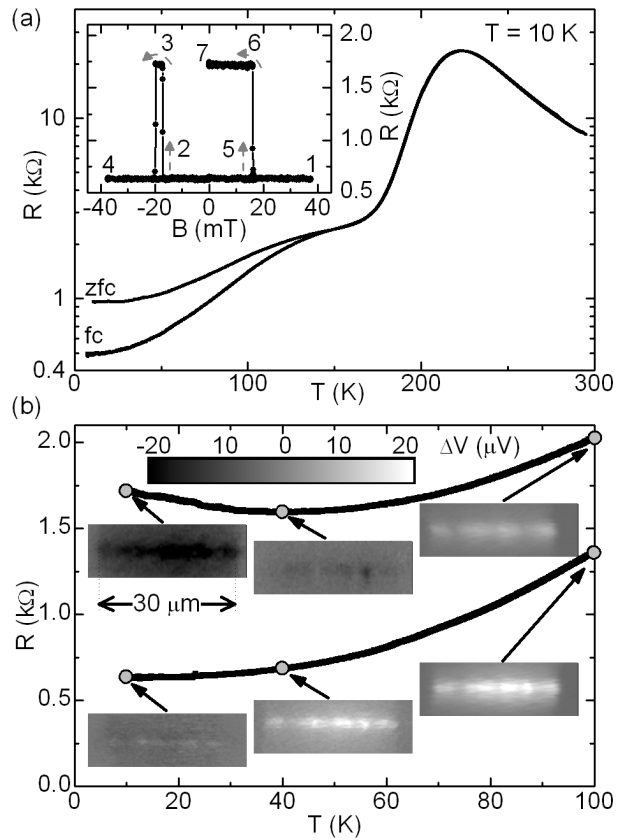}} %
\caption{TMR effect of LCMO bridge intersected by a
grain boundary ($I_b=5\,\mu\rm A$; $B$ is in-plane and
parallel to the GB). (a) $R(T)$ obtained by cooling in
zero field (zfc) and in $B=10\,\rm mT$ (fc). Inset:
$R(B)$ at $T=10\,\rm K$. $R$ switches between
antiparallel (high-resistance state) and parallel
(low-resistance state) magnetization of the LCMO films
adjacent to the GB by sweeping $B$ (from 1 to 7, as
indicated by dashed arrows). (b) $R(T)$ for high- and
low-resistance state at $B=0$. Insets show LTSLM
$\Delta V$ images for both states taken at different
$T$; upper inset shows $\Delta V$ scale.
\label{fig:2}}
\end{figure}

Fig.\ref{fig:2}(a) shows $R(T)$ in zero
field (zfc) and in $B=10\,\rm mT$ (fc).
At $T=220\,\rm K$ the film undergoes a
metal-to-insulator transition leading to
the well known high-field CMR\cite{Helmolt93,Jin94}. 
At $T<150\,\rm K$ the fc- and zfc-curves differ, \ie
the TMR of the GB appears. We attribute the
high-resistance state to antiparallel and the
low-resistance state to parallel orientation of the
magnetization of the LCMO films adjacent to the GB.
The relative orientation of the magnetization vectors
can be flipped between parallel and antiparallel by
sweeping $B$ at low $T$, as indicated in the inset of
Fig.\ref{fig:2}(a), showing switching between low- and
high-resistance state. Fig.\ref{fig:2}(b) shows the
$T$-dependence of the high- and low-resistance state
for $10\,{\rm K}<T<100\,\rm K$. Note that, at a given
temperature, $dR/dT$ differs for the two states. As
the LTSLM signal $\Delta V(x_0,y_0)$ changes sign with
$(\partial R/\partial T)|_{(x_0,y_0)}$ this
observation is consistent with the LTSLM images of the
GB [see insets of Fig.\ref{fig:2}(b)] taken for each
of the two states at three different temperatures.
Indeed, in all cases the $\Delta V$ signal obtained
along the GB turned out to be proportional to $dR/dT$,
which thus can be considered as the quantity we
primarily measure locally. In particular, at the
lowest temperature $T=10\,\rm K$, $dR/dT$ vanishes  in
the low-resistance state and is negative in the
high-resistance state
%
\footnote{This is due to the half-metallicity of the
doped manganites and the nature of the transport
processes across the disordered grain boundary region,
which have been discussed extensively in
Refs.~8,9,13,15,17.}.
%
Accordingly, the $\Delta V$
signal at the GB is vanishing in the low-resistance
state and is negative in the high-resistance state.
This fact allows discrimination by LTSLM imaging of
regions along the GB with parallel and antiparallel
orientation of the magnetization of the grains
separated by the GB, as will be discussed below.
Further note, that in each $\Delta V$ image, the
contrast varies along the GB, indicating that the GB
is not completely homogeneous. This can be attributed
to inhomogeneities (i) in the surface morphology,
causing a small variation of the absorbed laser
intensity, and (ii) in the microstructure of the film
close to the GB, which can induce inhomogeneities in
the tunnel barrier formed along the GB.

The $\Delta V$ images in
Fig.\ref{fig:2}(b) clearly show the
absence of any signal from the manganite
film (above and below the GB) with total
resistance $R_{film}$, consistent with
$(dR_{film}/dT)\ll(dR_{GB}/dT)$ at $T\le
100\,\rm K$. Hence, at low $T$, only
properties of the GB are probed by laser
imaging. This situation changes at higher
$T$, in particular close to $T_C$. In the
following, we demonstrate that LTSLM
allows discrimination of the
contributions from the film resistance
$R_{film}$ and the GB resistance $R_{GB}$
to the overall resistance
$R=R_{film}+R_{GB}$. From $\Delta V$
images taken at different $T$, we obtain
the evolution of $dR_{film}/dT$ and
$dR_{GB}/dT$ vs. $T$, and by integration
over $T$ we can reconstruct the local
$R_{film}(T)$ and $R_{GB}(T)$ curves.

\begin{figure}[tb]
\center{
\includegraphics[width=0.85\columnwidth,clip]
{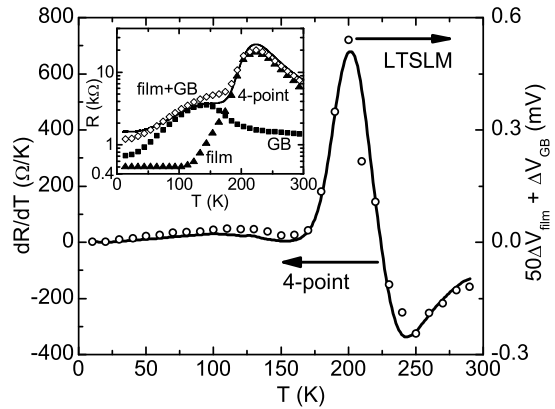}} %
\caption{$T$-dependence of LTSLM voltage
signals (open dots) and comparison with
integral $(dR/dT)(T)$ curve (solid line).
Inset: Local $R(T)$ curves, reconstructed
from integration of the LTSLM signals
over $T$. The spatial resolution of LTSLM
allows separation of the film (solid
triangles) and GB (solid squares)
resistances. The sum of both
contributions (open diamonds) gives
reasonably good agreement with the
overall resistance of the bridge (solid
line) measured by a conventional four
point technique. \label{fig:3}}
\end{figure}

For the following procedure the low
resistance state was chosen. From $\Delta
V$images we calculated the average
voltage signal $\Delta
V(y_0)=W^{-1}\int\Delta V(x_0,y_0)\,dx_0$
for linescans ($y_0$=const.) along the GB
($\Delta V^*_{GB}$) and along a
representative section of the bridge
($\Delta V_{film}$), $\sim20\,\mu\rm m$
above the GB. As the laser spot diameter
is much larger than the GB width (few
nm), the signal $\Delta V^*_{GB}$
contains a contribution from the film
(for $T>100\,\rm K$). To restore the
''pure'' GB signal for all values of $T$,
we thus take $\Delta V_{GB}=\Delta
V^*_{GB}-\Delta V_{film}$ to obtain
$(dR_{GB}/dT)(T)\propto\Delta V_{GB}(T)$
and $(dR_{film}/dT)(T)\propto\Delta
V_{film}(T)$. Due to the finite spot size
$d$, the signal $\Delta V_{film}$
corresponds to a fraction $d/L$ of the
bridge of length $L=70\,\mu\rm m$. Hence,
we expect the overall resistance $R$ and
the LTSLM signals to be related as
$(dR/dT)\propto\Delta V_{GB}+(L/d)\Delta
V_{film}\equiv\Delta V_{eff}$.
Fig.\ref{fig:3} shows $\Delta V_{eff}(T)$
as obtained from LTSLM images (with
$L/d=50$), and for comparison,
$(dR/dT)(T)$ as obtained from the
integral $R(T)$ curve. With an effective
spot size $d=L/50=1.4\,\mu\rm m$ we find
excellent agreement between both curves
over the whole temperature range. This
proves on the one hand that the LTSLM
signal is indeed of purely thermal
origin. On the other hand, the knowledge
of  the effective spot size allows
calculation of $\delta T(x_0,y_0)$, which
turns out to be $\approx 3\,\rm K$.

The inset of  Fig.\ref{fig:3} shows the local $R(T)$
curves, for both the film and the GB, as obtained from
integration of the LTSLM voltage signals over $T$. The
summation of both curves coincides reasonably well
with the integral $R(T)$ curve. Deviations are most
likely due to inhomogeneities in the bridge, which we
neglected in the analysis of the LTSLM signals. The
manganite film shows the well known metal-insulator
transition near $T_C$ ($\sim$220\,K), and a relatively
small (few $100\,\Omega)$ $T$-independent resistance
at $T\le 110\,\rm K$. In this low-$T$ regime, the GB
resistance clearly dominates the overall bridge
resistance and increases with $T$ up to a maximum
value of a few k$\Omega$ at $T\approx 140\,\rm K$,
approximately at the same temperature where the
low-field TMR effect disappears. This increase in
$R_{GB}$ with $T$ is consistent with a decreasing
magnetic order at the GB. Above 140\,K, however,
$R_{GB}$ decreases again with further increasing $T$,
which could be \eg attributed to a thermally activated
hopping mechanism dominating the transport across the
GB at higher $T$ in the paramagnetic regime.
Interestingly, the maximum in $R_{GB}$ appears at a
temperature which is close to the onset temperature of
increasing film resistance. A similar observation has
been reported in Ref.[6]. Finally, we note that the GB
resistance remains at a relatively high level even for
temperatures above $T_C$. This gives strong evidence
that the GB behaves as a tunnel barrier also in the
paramagnetic phase of the manganite film, suggesting a
tunneling transport mechanism.


\begin{figure}[tb]
\center{
\includegraphics
{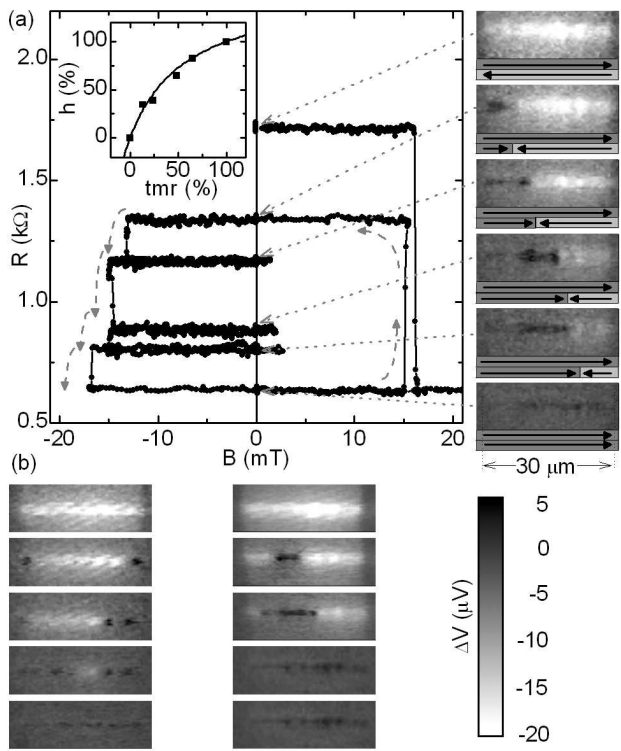}} %
\caption{Transition of LCMO GB junction
from high- to low-resistance state by
domain wall motion ($T$=10\,K;
$I_b=5\,\mu\rm A)$.
(a) $R(B)$ showing intermediate-resistance states
(dashed arrows indicate field sweep direction). LTSLM
images show separation of the GB into regions of high-
and low-resistance states (dotted arrows mark bias
points on $R(B)$). Consequently, one of the electrodes
must have developed a magnetic domain boundary which
nucleates at the left edge and moves along the GB to
the right edge, as $R$ is decreasing
(arrows indicate schematically magnetization
directions above and below GB). Inset: Fraction $h$ of
GB in high-resistance state vs relative TMR ratio ;
experiment (squares) and calculation (line).
(b) LTSLM images showing two different transitions
from the high- (top) to the low-resistance state
(bottom). Left column: domain wall nucleation at both
edges; right column: nucleation in the interior of the
bridge. \label{fig:4}}
\end{figure}
So far, we considered only the high- and
low-resistance states, with corresponding
resistances $R_{max}$ and $R_{min}$,
respectively. However,
intermediate-resistance states
($R_{min}<R<R_{max}$) can also be
stabilized by properly sweeping $B$, as
shown in Fig.\ref{fig:4}(a)
for $T$=10\,K.
The
corresponding
LTLSM images  show a striking change in
the contrast along the GB. One section
of normalized length $h\equiv H/W$ along
the GB is in the high-resistance state
($\Delta V<0$) while the other part is in
the low-resistance state ($\Delta
V\approx 0$).
Defining the relative TMR ratio as ${\rm
tmr}\equiv (R-R_{min})(R_{max}-R_{min})$
we can plot $h({\rm tmr})$ as obtained
from LTSLM images and $R(B)$ data
[c.f.~inset of Fig.\ref{fig:4}(a)]. From
our interpretation of the LTSLM signals
it follows that $h({\rm tmr})$ should
scale as $h({\rm tmr)} = {\rm tmr} \cdot
r_{max} /\{1+{\rm tmr}\cdot
(r_{max}-1)\}$, if we neglect $R_{film}$,
with $r_{max}\equiv R_{max}/R_{min}$.
This relation is also shown in the inset
of Fig.\ref{fig:4}(a) (as solid line).
The good agreement of experimental data
and calculated $h({\rm tmr})$ supports
our interpretation that LTSLM directly
probes the TMR.
The LTLSM images reveal the relative
orientation of the magnetization of the
LCMO films on both sides of the GB.
Hence, one of the LCMO electrodes must
have developed a domain boundary where
the magnetization flips.
%
However, we do
not know how far the domain walls
continue into the LCMO films. Such
complementary information could be \eg
obtained by Kerr microscopy, which,
however, requires smooth and highly
reflective surfaces in contrast to LTSLM.
How reproducible are the above results?
We performed measurements as in
Fig.\ref{fig:4} for two samples and about
20 times, also at various $T$. In most
cases we found nucleation of  a domain
wall at one edge of the GB which then
proceeds towards the other edge [\cf
Fig.\ref{fig:4}(a)]. For the sample
discussed here, in seven out of ten
measurements at 10\,K we found domain
wall nucleation at one edge. In two cases
we found domain wall nucleation at both
edges [\cf Fig.\ref{fig:4}(b), left
column], and only in one case a domain
nucleated in the interior of the bridge
[Fig.\ref{fig:4}(b), right column]. This
process occurs more rarely, because it
requires simultaneous creation of two
domain walls, which is energetically less
favorable than producing seed domains at
the edges of the GB.


In conclusion, we demonstrated spatially
resolved detection of the TMR of LCMO
grain boundaries by LTSLM. We observe on
one hand fairly homogeneous low- and
high-resistance states, consistent with
fully parallel and antiparallel
orientation of  the magnetization of the
LCMO electrodes at the GB, respectively.
As LTSLM measures locally $dR/dT$, it
allows to reconstruct and compare the $T$
dependence of both, the manganite film
and GB resistance within a single bridge
containing a GB. Hence, LTSLM, if
combined with integral magnetotransport
measurements, may provide important
information for the understanding of
low-field TMR effects in manganites.
Finally, we have also detected nontrivial
intermediate states where the
magnetization inside of one of the LCMO
films has flipped. For such situations,
LTLSM is a powerful tool, as it directly
images the quantity of interest
(magnetoresistance) and its dependence on
magnetic domain formation, which is
important for any MR element. As LTSLM
does not rely on smooth or reflective
surfaces, this technique may thus provide
important insight into the detailed
behavior not only of manganite grain
boundaries, but also of many other
magnetoresistive devices.

We thank M. Peschka, R. P. Huebener, M. Fischer, P.
M\"{u}hlschlegel and T. Schwarz for their contributions to
the development of the laser microscope. This work was
supported by the Deutsche Forschungsgemeinschaft.

\bibliography{\bibdir xmr}

\end{document}